\documentclass{ecai}
\usepackage{graphicx}
\usepackage{latexsym}
\usepackage{graphicx}
\usepackage{subcaption}
\usepackage{subfloat}
\usepackage{stfloats}
\usepackage{algorithm}
\usepackage{algorithmic}
\usepackage{bbm}
\usepackage{amsmath}
\usepackage{romannum}
\usepackage{mathtools}

\def \ALGO{MESE\ }
\renewcommand{\thefootnote}{\fnsymbol{footnote}}

\ecaisubmission   

\begin{document}

\begin{frontmatter}

\title{Multi-agent Exploration with Sub-state Entropy Estimation}

\author[\textdagger]{\fnms{Jian}~\snm{Tao}\thanks{Corresponding Author. Email: tj22@mails.tsinghua.edu.cn}}
\author[\textdagger]{\fnms{Yang}~\snm{Zhang}}
\author[]{\fnms{Yangkun}~\snm{Chen}} 
\author[]{\fnms{Xiu}~\snm{Li}}

\address[]{Shenzhen International Graduate School\\
Tsinghua University}
\address[\textdagger]{Equal Contribution}


\begin{abstract} Researchers have integrated exploration techniques into multi-agent reinforcement learning (MARL) algorithms, drawing on their remarkable success in deep reinforcement learning. Nonetheless, exploration in MARL presents a more substantial challenge, as agents need to coordinate their efforts in order to achieve comprehensive state coverage. Reaching a unanimous agreement on which kinds of states warrant exploring can be a struggle for agents in this context. We introduce \textbf{M}ulti-agent \textbf{E}xploration based on \textbf{S}ub-state \textbf{E}ntropy (\ALGO) to address this limitation. This novel approach incentivizes agents to explore states cooperatively by directing them to achieve consensus via an extra team reward. Calculating the additional reward is based on the novelty of the current sub-state that merits cooperative exploration. \ALGO employs a conditioned entropy approach to select the sub-state, using particle-based entropy estimation to calculate the entropy. \ALGO is a plug-and-play module that can be seamlessly integrated into most existing MARL algorithms, which makes it a highly effective tool for reinforcement learning. Our experiments demonstrate that \ALGO can substantially improve the MAPPO's performance on various tasks in the StarCraft multi-agent challenge (SMAC).
\end{abstract}

\end{frontmatter}

\section{Introduction}
\label{sec_intro}

\begin{figure*}[tp]
	\centering
    
	\begin{subfigure}{0.33\linewidth}
		\centering
		\includegraphics[width=0.95\linewidth]{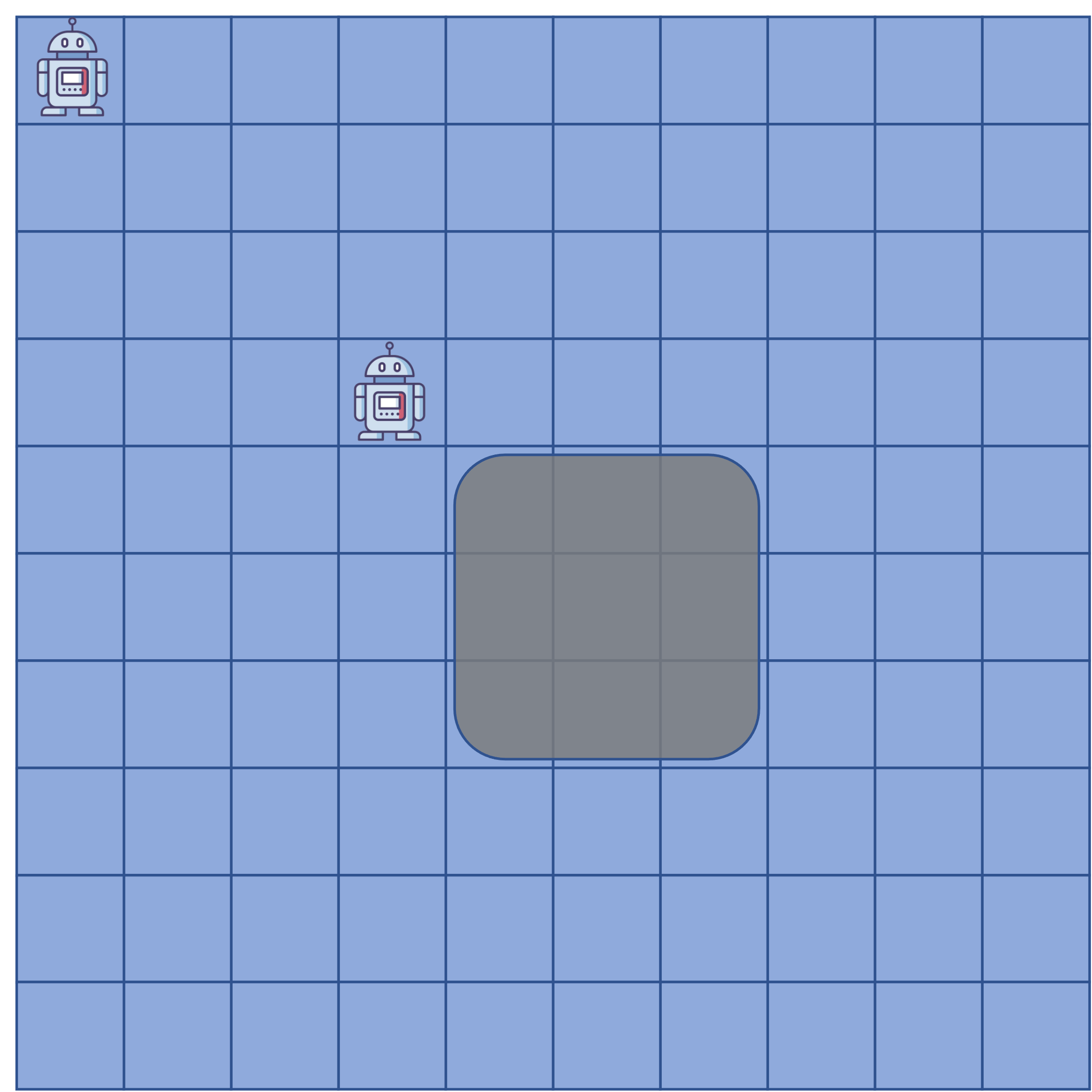}
	\end{subfigure}\hspace{20mm}
	\centering
	\begin{subfigure}{0.33\linewidth}
		\centering		\includegraphics[width=0.95\linewidth, height=0.95\linewidth]{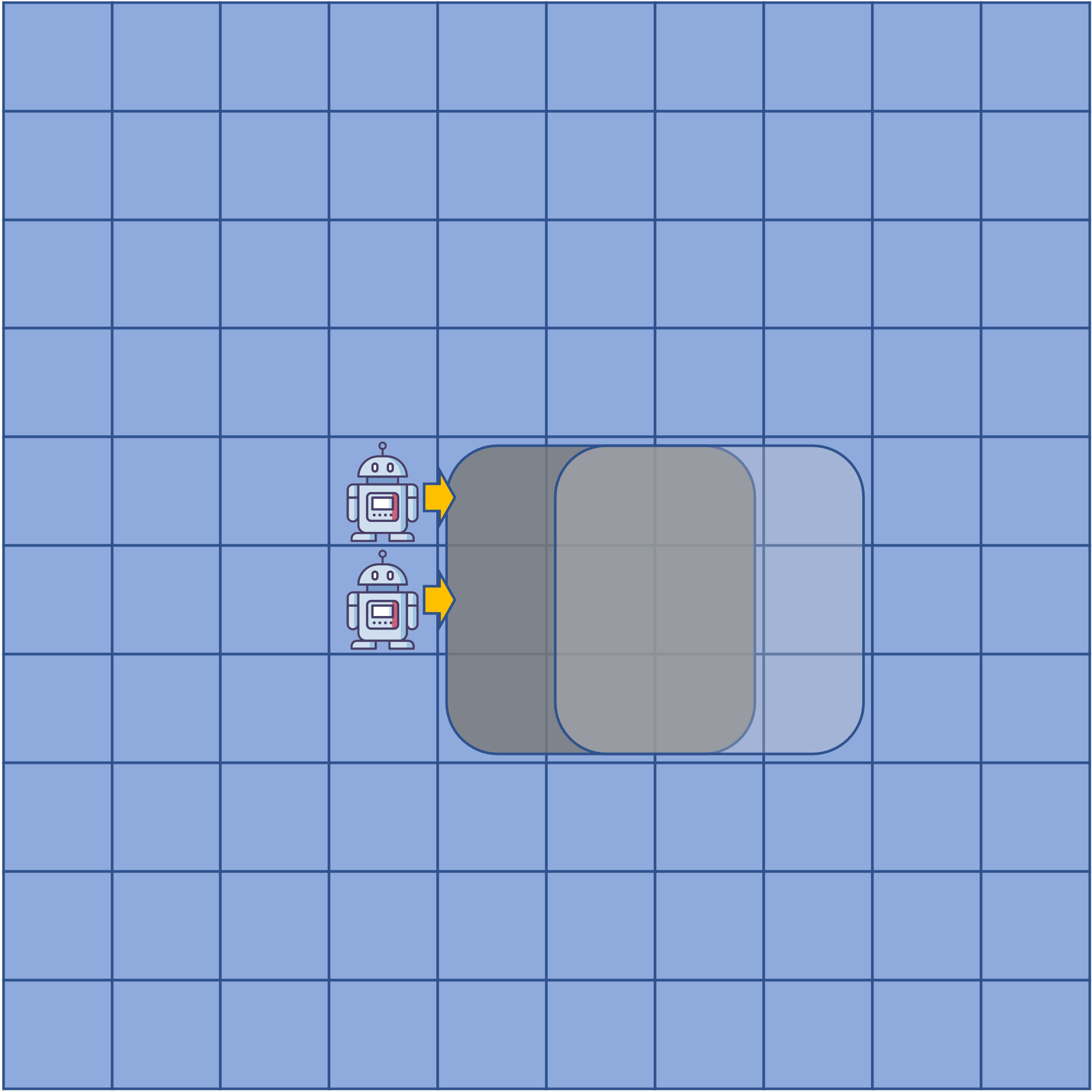}
	\end{subfigure}
	\caption{Pushing Box is a toy game that requires exploration. In this game, agents can move in four directions, and the state includes the coordinates of all entities. Agents receive a shared reward when the box is pushed to the destination and the game ends simultaneously. The box can only be pushed if (1) all agents are on the same side of the box and are adjacent to the box and (2) all agents move to the other side of the box simultaneously.}
    \label{fig:pushbox}
\end{figure*}

Reinforcement Learning (RL) and Multi-Agent Reinforcement Learning (MARL) have demonstrated remarkable success in decision-making, control, and planning, spanning diverse domains such as video games \cite{ref_atari}, Go \cite{ref_go}, robot control \cite{ref_robot}, and MOBA games \cite{ref_moba}. Moreover, RL has emerged as a key player in natural language processing, exemplified by its use in fine-tuning the ChatGPT\cite{gpt} language model for tasks such as text completion and question answering. The potential of RL is immense and continues to expand as research in multi-agent systems advances in tandem with RL.

However, multi-agent reinforcement learning encounters more significant challenges than single-agent reinforcement learning. Several factors contribute to these challenges:

The first factor contributing to the challenges is the exponential expansion of the state-action space's magnitude. Compared to single-agent reinforcement learning, the joint state-action space in multi-agent reinforcement learning increases exponentially with the number of agents. This exponential increase in complexity makes exploring the environment in MARL more difficult. Therefore, a practical exploration strategy is crucial for MARL algorithms to reduce the cost of training.

The second challenge of multi-agent reinforcement learning is coordinated exploration, a tricky problem due to partial observation and non-stationary issues in MARL. Since each agent's policy influences the environmental transition, continuous updating, and synchronous decision-making introduce inherent randomness into the system. Thus, exploring without purpose often fails in tasks that require agents to cooperate. Figure \ref{fig:pushbox} illustrates a typical environment heavily requiring coordinated exploration. In the game \textit{PushBox}, agents must cooperate to push the box to the goal position. The box will move if both agents move in the same direction only. The agents receive a shared reward when the box reaches the destination. Although this task seems simple, it is instead significantly challenging. The likelihood of achieving the correct state-action pair (both agents positioned on the same side of the box and moving towards the opposite side) is remarkably low without task-specific demonstration or handcrafted heuristic guidance.

Intuitively, an increased visitation frequency of the critical state-action pairs mentioned above can significantly enhance sample efficiency. However, current exploration methods do not prioritize these critical states and only perform general, non-cooperative exploration. Therefore, we aim to investigate ways to enhance the probability of encountering these pivotal states, which can promote more efficient and cooperative exploration. 

Researchers have preliminarily classified existing exploration methods in multi-agent reinforcement learning into two major categories. The first category is uncertainty-oriented exploration, which originates from the principle of Optimism in the Face of Uncertainty (OFU). The key idea of this category is to estimate the adequacy of exploration with the randomness of value estimates. The second category is intrinsic motivation-oriented exploration \cite{ref_exp_cate}. An intuitive explanation for this category is that children often learn about the world without external rewards but with intrinsic rewards such as curiosity or an adventurous spirit. These methods typically use a designed intrinsic reward to measure the novelty of states for reshaping real rewards from the environment and guiding agents to explore. Both methods employ a design exploration paradigm based on a particular property of encountered states. However, such a design is not optimal as it relies solely on \textit{encountering useful states} rather than leveraging previous experience to determine which states are worth exploring. This approach can result in instability and inefficiency in exploration, especially given the low probability of encountering crucial and critical states. For example in Figure \ref{fig:pushbox}, the exploration algorithm may not recognize the importance of the box coordinates until the box moves, which occurs with low probability. However, as our task involves moving a box, it is evident that essential information underlies the box coordinates before any agent attempts to move the box. Furthermore, while exploration helps encourage agents to move independently, it may not necessarily lead to effective cooperative behavior.

By measuring the diversity of a specific part of the state, we can recognize the importance of box coordinates and develop a more efficient exploration policy. We can consider this part of the state as a multidimensional continuous/discrete random variable and quantify its diversity by calculating the corresponding random variable's entropy. To calculate the intrinsic reward, a sub-state (a specific part of the whole state) can be selected as input for a module, such as a random network distillation (RND). The advantage of this approach is that exploration becomes more efficient and more directional as it no longer solely depends on encountering unknown states. Moreover, this state processing reduces the impact of dimensions with high entropy, which are easily explored, on intrinsic rewards. Consequently, the agent is encouraged to focus on the essential parts and explore more effectively.  

To implement our method, we propose \textbf{M}ulti-agent \textbf{E}xploration based on \textbf{S}ub-state \textbf{E}ntropy (MESE). As our approach builds on estimating the importance of a specific part of the state, a module to compute intrinsic reward based on the novelty of the state is needed. For simplicity, we directly adopt Random Network Distillation (RND) \cite{rnd} to compute intrinsic reward. Consequently, our approach integrates RND with an entropy-driven subspace search module. The method allows us to accurately capture the diversity of a particular part of the state, which is crucial for designing effective exploration policies. Meanwhile, it can be easily incorporated into existing multi-agent reinforcement learning frameworks without extensive modifications. We also utilize a non-parametric particle-based entropy estimator to calculate the differential entropy for continuous variables, which ensures an unbiased estimate of entropy. This approach enables us to extract and leverage the critical information in the state and identify the dimensions that are most likely to lead to helpful exploration. Overall, our method provides a more efficient way to explore unknown states and improves the stability of the multi-agent reinforcement learning process. Last but not least, we claim that the main contribution of our approach is introducing an entropy-based subspace search module to guide intrinsic motivation-oriented exploration towards coordinated exploration for multi-agents, which implies that RND only serves as a module for computing intrinsic reward and can be directly replaced with any similar module utilizing the novelty of state to obtain intrinsic motivation.

Our contributions are as follows:
\begin{itemize}
    \item[a)] We propose a new plug-and-play module, Multi-Agent Exploration based Sub-state Entropy(MESE), to identify critical states that are highly resistant to change, enhancing agents' ability to explore their environment effectively.
    \item[b)] We analyze the difficulties encountered in multi-agent reinforcement learning: exponential expansion of the state-action space's magnitude and lack of coordinated exploration, which lead to slow convergence of the estimated value function and diminish sample efficiency. 


    \item[c)] We integrate our approach into MAPPO\cite{ref_MAPPO}, and the results demonstrate an improvement in exploration efficiency with the updated algorithm. It dramatically improves the convergence rate in the benchmark StarCraft multi-agent challenge (SMAC) with precisely the same parameters and network structure.
\end{itemize}



\section{Background and Related work}
\subsection{Cooperative Multi-Agent Reinforcement Learning}

Cooperative multi-agent reinforcement learning task can be modeled as Decentralized Partially Observable Markov Decision Process (Dec-POMDP). In this setting, each agent takes action independently with its own policy, shares a team reward and collaborates with each other to maximize team reward. It is usually described as a tuple as $(\mathcal{A}, S, O, U, P, r, \gamma, \rho_0)$. $\mathcal{A}=\{1, 2, \cdots, n\}$ means the set of $n$ agents. Denote global state space as $S$ and the observation space of the agents as $O=\{O_1, O_2, \cdots, O_n\}$ respectively. And let $U=\{U_1, U_2, \cdots, U_n\}$ denote the action space of the agents. The agents take a joint action $\mathbf{u}_t=\{u_t^1, u_t^2, \cdots, u_t^n\}$ according to observations and state at each timestep $t$ to interact with the environment and obtain the next state with the state transition function $P(s_{t+1}|s_t,\mathbf{u}_t): S\times U\times S\rightarrow[0, 1]$. $r(s_t, \mathbf{u}_t):S\times U\rightarrow\mathbbm{R}$ denotes the team reward function from the environment. $\gamma\in [0, 1)$ is a discount factor and $\rho_0:S\rightarrow\mathbbm{R}$ is the distribution of the initial state $s_0$. The goal of cooperative multi-agent reinforcement learning is to maximize the expectation of discounted cumulative return $J(\boldsymbol{\pi})=\mathbbm{E}_{s_0, \mathbf{u}_0, r_0, \cdots}[\sum_{t=0}^\infty\gamma^t r_{t}]$, where $\boldsymbol{\pi}=\{\pi_1, \pi_2, \cdots, \pi_n\}$ means joint policy.

\begin{figure*}[tp]
    \centering
    \includegraphics[width=0.8\linewidth]{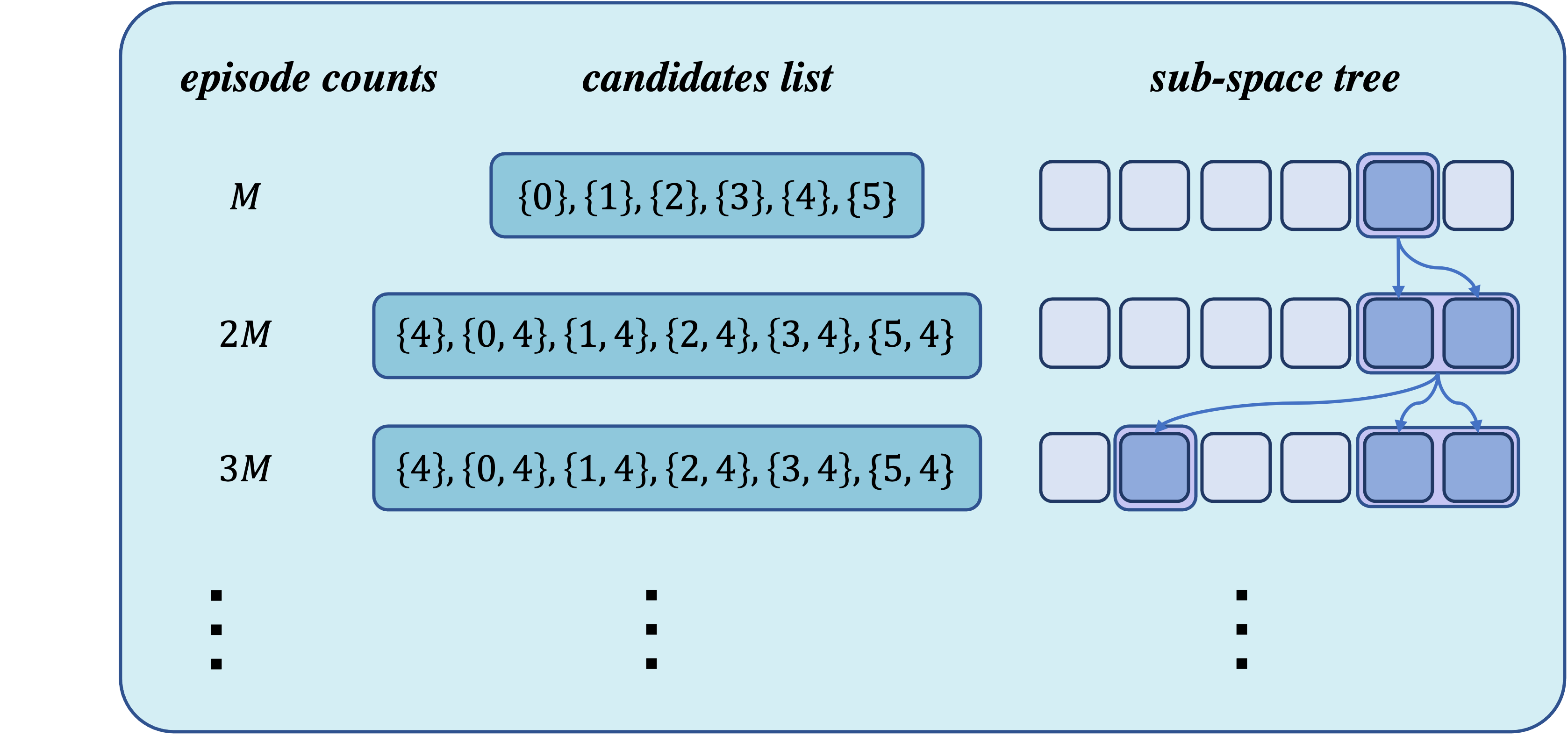}
    \caption{sub-space tree expansion}
    \label{fig:subspace_tree}
\end{figure*}
\subsection{Exploration in RL and MARL}
In this section, we introduce previous research related to exploration in both single-agent RL and multi-agent RL.

\subsubsection{Exploration in Single Agent Reinforcement Learning} 
$\epsilon$-greedy is the most basic exploration method in RL\cite{ref_DQN}: the agent chooses the action greedily with a probability of $1-\epsilon$ and a random action with a probability of $\epsilon$. It is also common to use Gaussian noise in deterministic policy networks like DDPG\cite{ref_DDPG}. Besides Gaussian noise (white noise), other noises like Ornstein-Uhlenbeck (OU) noise\cite{ou_noise} and colored noise\cite{colored_noise}\cite{ref_pink} have been shown to have different effects on exploration. In algorithms where the policy is stochastic like SAC\cite{ref_SAC} or MPO\cite{ref_MPO}, the action sampling itself introduces randomness. Another exploration strategy that is closer to our work is intrinsic motivation-oriented exploration, which encourages the agent to explore with an extra bonus. Methods for evaluating the novelty of a state by counting or neural network constitute a subcategory of these works. These techniques are geared towards encouraging agents to travel to states they've never or rarely visited. The count-based methods establish the intrinsic reward as the inverse proportion of the visit counts of state $N(s_t): R_{in}(s_t)=1/N(s_t)$. In the absence of a special design, it is typically challenging to measure counts in a large or continuous state space\cite{trpo_hash}\cite{a3c+}\cite{density_model}. Random Network Distillation (RND)\cite{rnd} estimates the state novelty by distilling a fixed random network into the other trainable network. The intrinsic reward is the prediction error of predicting features generated by the two networks. The key idea is that the prediction error is small if some state is frequently visited. The reshaped rewards policy used in training is greater if the next states are never visited, because the less common the states, the greater the prediction error of RND. The way processing states or features using neural networks can deal with both discrete and continuous space naturally.

\subsubsection{Exploration in Multi-agent Reinforcement Learning} 
Learning Individual Intrinsic Reward (LIIR)\cite{liir}, which learns an extra proxy critic for each agent, extends a similar idea in single-agent domains, with the input of intrinsic rewards and extrinsic rewards. \cite{cmae} define the intrinsic reward function with states in sub-space rather than the whole state space, which is used for more purposeful collaborative exploration. However, the intrinsic reward in \cite{cmae} is count-based so it does not handle continuous spaces well. \cite{EMAX} trains ensembles of value functions for each agent to resolve the most important challenges of exploration and non-stationarity. Nevertheless, because it employs numerous ensemble models with significant computational costs, our model requires only one additional pair of encoders, which is more affordable.


\section{Methodology}
In this section, we formally propose the MESE method. We first give intuitive demonstrations of our methods based on the distinguishments between global state and sub-state, then we describe how to estimate differential entropy via particle-based estimator in Section \ref{sec:estimator} and how to choose subspace to calculate intrinsic reward in Section \ref{sec:subspace} respectively.

\subsection{Motivation}\label{sec:motivation}
Standard intrinsic motivation-oriented exploration methods ignore a fact: some parts of the state vary less than other parts. For instance, in Montezuma's Revenge, a classic reinforcement learning study environment, the state information on the existence or absence of a key on the map is more significant than the state of the area where the agent is situated, and it mostly stays unchanged.
As a result, the probabilities of two states, $s$ and $s'$, are different. However, both states are worth exploring equally, i.e., a mechanism to encourage exploration by differentiating subspace variation diversity is needed, which encourages agents to explore according to state subspace in which the sub-state has less diversity instead of the diversity of the entire state space. More importantly, using sub-state instead of complete state space avoids the influence of other dimensions on intrinsic reward and makes exploration more directional.

We select sub-state according to differential Shannon entropy
\begin{equation}
    H(\boldsymbol{x})=-\mathbbm{E}(\ln{p(\boldsymbol{x})})=-\int p(\boldsymbol{x})\ln{p(\boldsymbol{x})} d\boldsymbol{x},\label{diff_ent}
\end{equation}
an extension to continuous probability distributions, where $\boldsymbol{x}$ is a multidimensional continuous random variable with the probability density function $p(\boldsymbol{x})$, $\boldsymbol{x}\in\Omega$, $\Omega \subset \mathbbm{R}^{n_{\boldsymbol{x}}}$. 
\begin{figure*}[t]
    \centering
    \includegraphics[width=0.7\linewidth]{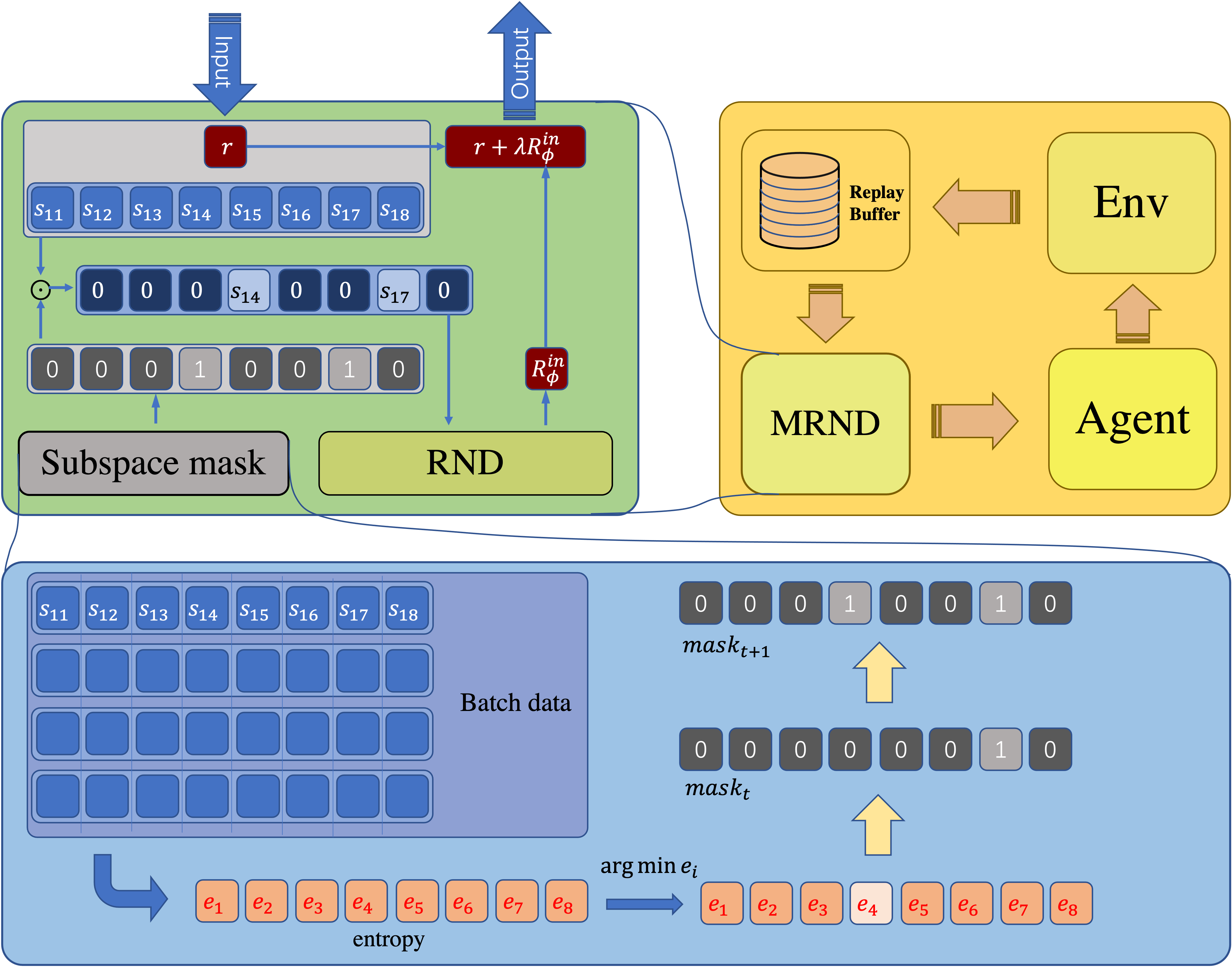}
    \caption{Schematic for \ALGO\ \textbf{Right}: It is a usual multi-agent reinforcement learning flowchart. The agents interact with the environment and obtain rewards from the environment until finishing the game. The episode (transition) is stored in the replay buffer for training the agents' policy. The MRND is a plug-and-play module used for the episode (transition) from the buffer before updating parameters of agents.\ \textbf{Left}: Left figure is the method of how \ALGO updates the subspace (mask) per $M$ episode. \textbf{Middle}: Flowchart of our plug-and-play module (use a transition as a demonstration). The input is $r_t$ and $s_{t+1}$ of the transition. \ALGO values novelty of $s_{t+1}$ with a masked state.}  
    \label{fig:mrnd}
\end{figure*}

\subsection{Particle Differential Entropy Estimator}\label{sec:estimator} 
In this section, we introduce the entropy calculation method illustrated in Figure \ref{fig:mrnd}(Left) (the part from states to entropy), which is fundamental for generating a sub-space mask.

We use the $k$-th nearest neighbor entropy estimator\cite{ref_Differential_entropy_estimation} to estimate the differential entropy. It is one of the non-parametric approaches studied in statistical literature for decades. This method calculates the distribution diversity with the distance between each sampled point and its $k$-th neighbor point. Specifically, let $Z={z_i|i=1,2,\cdots, N}$ be a sampled data points set from an unknown continuous distribution $\pi(\boldsymbol{x})$, the estimation of entropy is 
\begin{equation}\label{equ_estimate}
    \hat{H}(\pi)=-\frac{1}{N}\sum^N_{i=1}\ln{\frac{k}{N}\frac{\Gamma(n_x/2+1)}{{\lVert\boldsymbol{x}^i-\boldsymbol{x}^i_{k-NN}\rVert}^{{n_x}\pi^{n_x/2}}}}+\ln{k}-\Psi(k),
\end{equation}
where $\Gamma(x)$ is the Gamma function and $\Psi(x)=\frac{\mathrm{d}}{\mathrm{d}x}\ln{\Gamma(x)}$ is the digamma function. $k$ is a hyperparameter as a part of the bias correction term. $\boldsymbol{x}_i^{k-NN}$ is $k$-th nearest neighbors. For a $n_x$ dimensional sphere with radius $\rho$, the volume is 
\begin{equation}\label{equ_volume}
    V(\rho,n_x)=\frac{\rho^{n_x}\pi^{n_x/2}}{\Gamma(n_x/2+1)},
\end{equation}
we can substitute equation (\ref{equ_volume}) into equation (\ref{equ_estimate}) to obtain
\begin{equation}
    \hat{H}(\pi)\propto \sum_{i=1}^N\ln{V_i^k},
\end{equation}
where $V_i^k$ is the volume of the hypersphere of radius $\lVert\boldsymbol{x}_i-\boldsymbol{x}^i_{k-NN}\rVert$. To simplify the calculation, \ALGO only uses the sum of the log of the average value of Euclidean distance between each point and its all $k$ nearest neighbor. The average can improve the accuracy of estimation.
\begin{equation}
    \hat{H}_{\ALGO}(\pi)\coloneqq\sum_{i=1}^N\ln{\frac{1}{k}\sum_{\boldsymbol{x}_i^j\in N_k(\boldsymbol{x}_i)}\lVert\boldsymbol{x}_i-\boldsymbol{x}_i^j\rVert},
\end{equation}
$N_k(\boldsymbol{x}_i)$ is the set including $k$ nearest neighbors of sampled point $\boldsymbol{x}_i$.


\begin{figure*}[tp]
	\centering
	\begin{subfigure}{0.33\linewidth}
		\centering
		\includegraphics[width=0.95\linewidth]{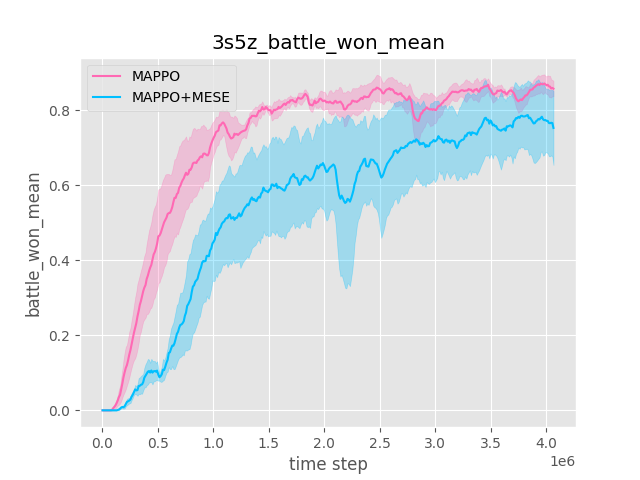}
		\caption{3s5z}
		\label{3s5z}
	\end{subfigure}
	\centering
	\begin{subfigure}{0.33\linewidth}
		\centering
		\includegraphics[width=0.95\linewidth]{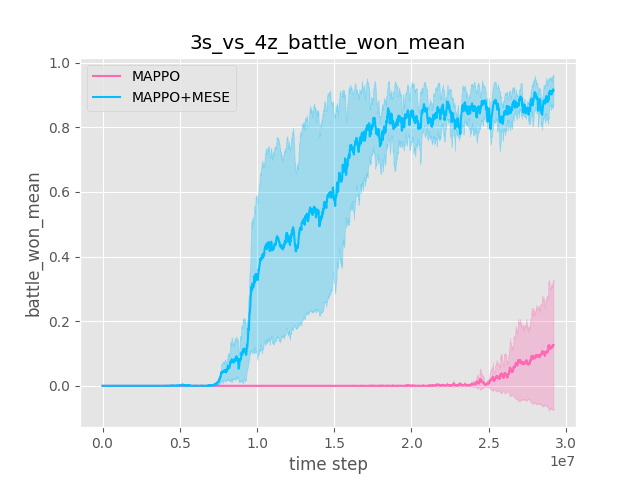}
		\caption{3s vs 4z}
		\label{3s_vs_4z}
	\end{subfigure}
	\centering
	\begin{subfigure}{0.33\linewidth}
		\centering
		\includegraphics[width=0.95\linewidth]{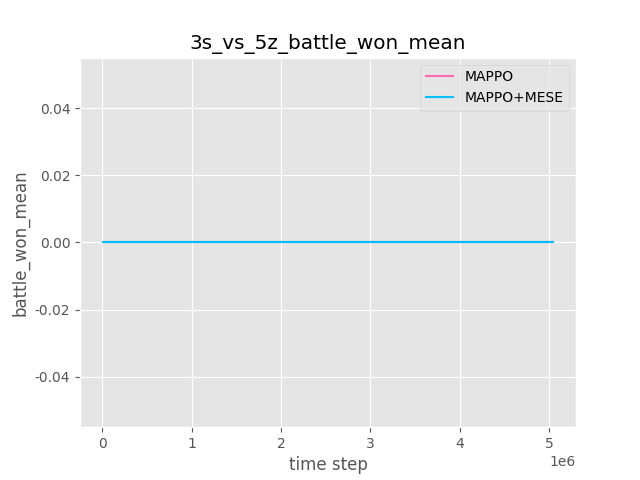}
		\caption{3s vs 5z}
		\label{3s_vs_5z}
	\end{subfigure}

 \quad
 
        \begin{subfigure}{0.33\linewidth}
		\centering
		\includegraphics[width=0.95\linewidth]{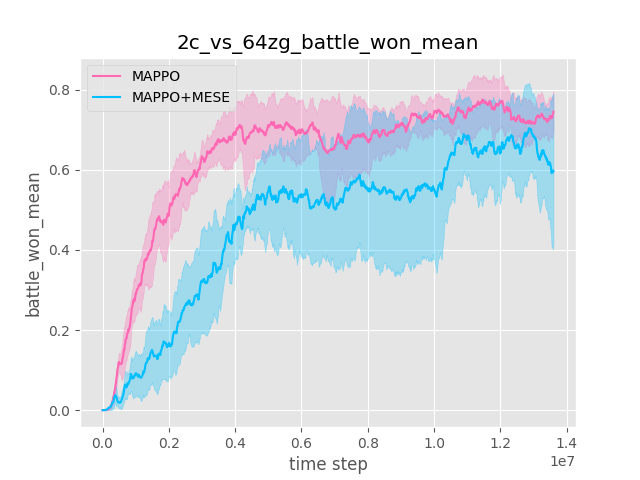}
		\caption{2c vs 64zg}
		\label{2c_vs_64zg}
	\end{subfigure}
	\centering
	\begin{subfigure}{0.33\linewidth}
		\centering
		\includegraphics[width=0.95\linewidth]{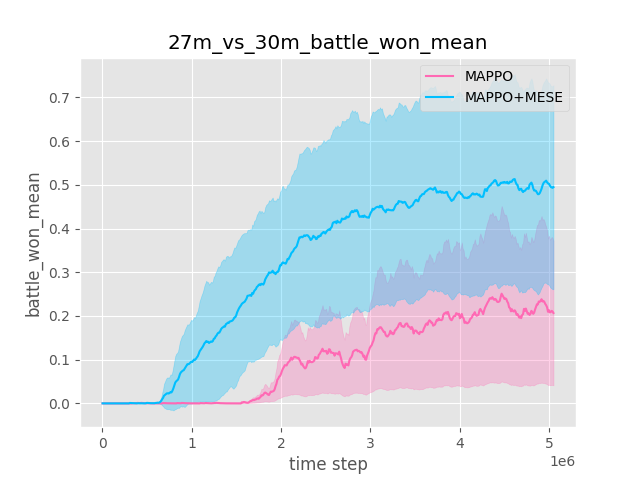}
		\caption{27m vs 30m}
		\label{27m_vs_30m}
	\end{subfigure}
	\centering
	\begin{subfigure}{0.33\linewidth}
		\centering
		\includegraphics[width=0.95\linewidth]{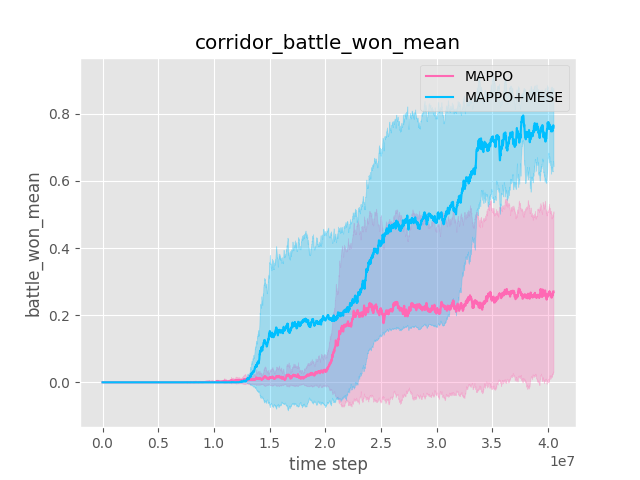}
		\caption{corridor}
		\label{corridor}
	\end{subfigure}
        
	\caption{StarCraft 2 win rate}
\label{fig_sc2_1}
\end{figure*}

\subsection{Sub-state Selection Based on Uncertainty}\label{sec:subspace}
In this section, we introduce how to choose a sub-state as input of an intrinsic reward function given a specific uncertainty function.

Given an uncertainty function $D(s_{sub}):\mathbbm{R}^{n_s}\rightarrow\mathbbm{R}$, whose input is a sub-state $s_{sub}$, we choose a sub-state from several candidates based on it. \ALGO generate the candidates from a subspace tree expansion process, which means we explore a low-dimensional subspace first and expand it gradually because the number of the whole subset (the size of the power set) is too large to calculate all of them. The sub-state is defined as follows:
\begin{equation}
    s_I=\mathcal{M}_I(s),
\end{equation}
where $\mathcal{M}_{I}(s)$ is a mapping that masks certain dimensions of the state space, resulting in a sub-state that only retains the dimensions specified by the index set $I\subseteq J=\{1,2,\dots,n\}$. To illustrate, suppose $n=3$ and $J=\{1,2,3\}$ represents the full state space. If we choose $I=\{1,2\}$, the resulting sub-state $s_I$ would only retain the first two dimensions of $s$ while masking the third. This process allows us to focus on a smaller subset of relevant dimensions to our analysis while ignoring the others. 

Specifically, we choose the sub-state from one dimension to $n$ dimensions. We change the sub-state per $M$ training, where $M$ is a hyperparameter to control the frequency of updating the subspace. The initial one-dimensional subspace, also the root of the subspace tree, is determined according to the diversity of the corresponding dimension of the state. The smaller the corresponding one-dimensional sub-state diversity, the more agents must explore cooperatively. \ALGO can compute a new subspace iteratively once determining the initial subspace. The covariance can express diversity, but for non-gaussian densities and especially for multimodal densities, the covariance can be misleading\cite{ref_Differential_entropy_estimation}. In \ALGO, we use conditional differential entropy introduced in Section \ref{sec:estimator} to measure the diversity of a subspace as follows:
\begin{equation}\label{equ_cal_d}
    d_j=D(\mathcal{M}_{\boldsymbol{C}_j}(\boldsymbol{B})),
\end{equation}
where $D$ denotes the diversity calculator (entropy estimator here), $\boldsymbol{C}_j$ is $j$-th candidate in current candidate list $\boldsymbol{C}$ and $\boldsymbol{B}$ is the batch data used for calculation. 

Specifically, given a selected subspace $S_\mathcal{M_I}\in S$, we define the new node set added to the subspace tree as follows:
\begin{equation}\label{equ_add_node}
    I_{expand}=\{I_e=I\cup i\ |\ i\in I^-\},
\end{equation}
where $I^-$ is the complement of $I$. Furthermore, we define the candidate list as the union of $I_{expand}$ and the current subspace $I$. Based on this, \ALGO expands the index set $I$ with minimum conditional entropy in the candidate list consisting of the lowest leaf node from the expanded subspace tree:
\begin{equation}
    I'\leftarrow I\cup \{j\}, j=\arg\min H(s_j|s_I),
\end{equation}

We summarize the update process in Algorithm \ref{algo_subspace} and illustrate the sub-space tree expansion in Figure \ref{fig:subspace_tree}.

\begin{algorithm}
    \renewcommand{\algorithmicrequire}{\textbf{Input:}}
    \caption{sub-space tree expansion}
    \label{algo_subspace}
    \begin{algorithmic}[1]
        \REQUIRE diversity function $D$, candidate list $\boldsymbol{C}$, mask index $\boldsymbol{I}$, batch data $\boldsymbol{B}$, episode count $e$, update interval $M$
        \STATE calculate diversity for each candidate sub-space based on (\ref{equ_cal_d})
        \STATE choose the candidate with minimal diversity $C_{min}$ as new mask index $I'$
        \IF{$e$ mod $M$ = 0}{
                \STATE update candidate list with $I_{expand}$ in (\ref{equ_add_node})
            } 
        \ELSE \STATE update candidate list with $I_{expand}\cup \boldsymbol{I}$
        \ENDIF
    \end{algorithmic}
\end{algorithm}

\begin{figure*}[tp]
	\centering
	\begin{subfigure}{0.33\linewidth}
		\centering
		\includegraphics[width=0.95\linewidth]{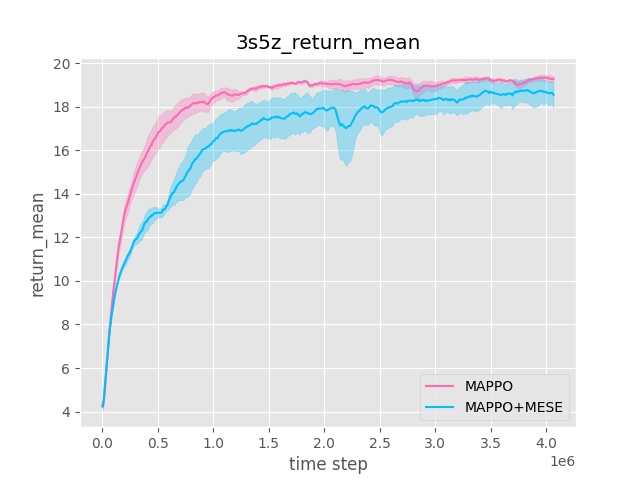}
		\caption{3s5z}
		\label{3s5z}
	\end{subfigure}
	\centering
	\begin{subfigure}{0.33\linewidth}
		\centering
		\includegraphics[width=0.95\linewidth]{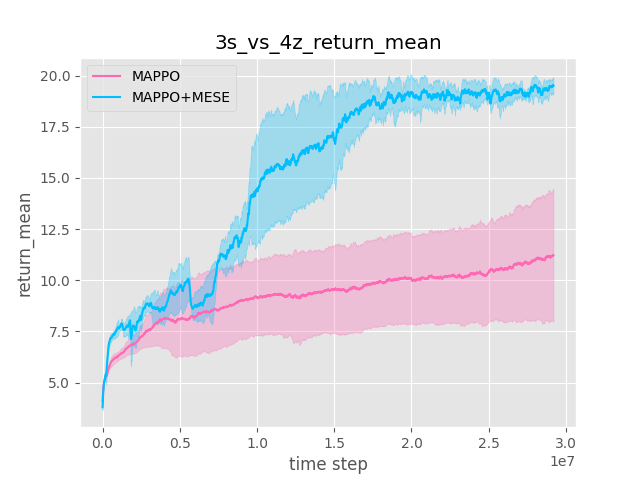}
		\caption{3s vs 4z}
		\label{3s_vs_4z}
	\end{subfigure}
	\centering
	\begin{subfigure}{0.33\linewidth}
		\centering
		\includegraphics[width=0.95\linewidth]{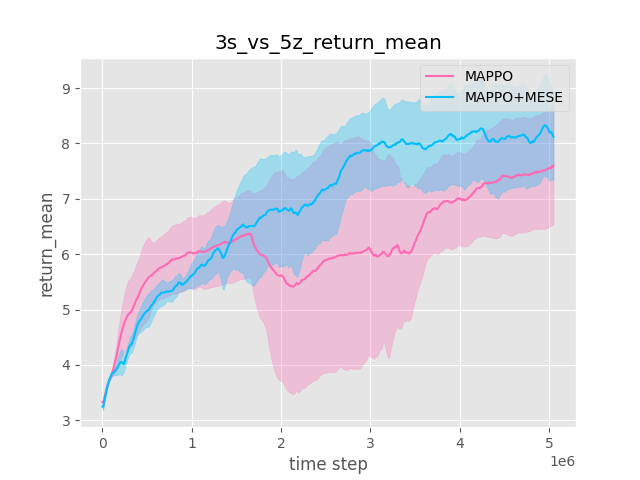}
		\caption{3s vs 5z}
		\label{3s_vs_5z}
	\end{subfigure}

 \quad
 
        \begin{subfigure}{0.33\linewidth}
		\centering
		\includegraphics[width=0.95\linewidth]{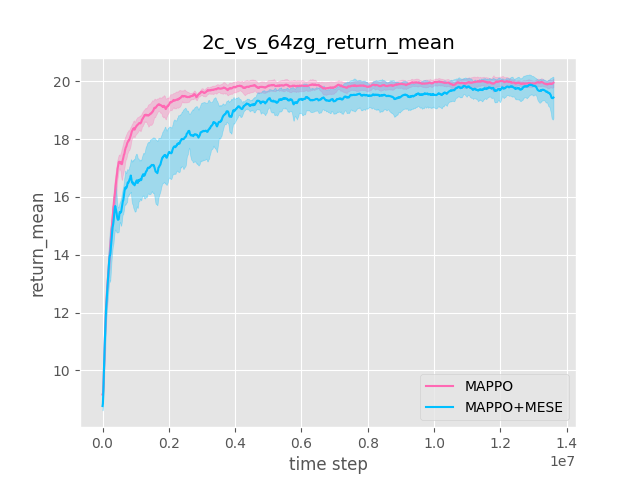}
		\caption{2c vs 64zg}
		\label{2c_vs_64zg}
	\end{subfigure}
	\centering
	\begin{subfigure}{0.33\linewidth}
		\centering
		\includegraphics[width=0.95\linewidth]{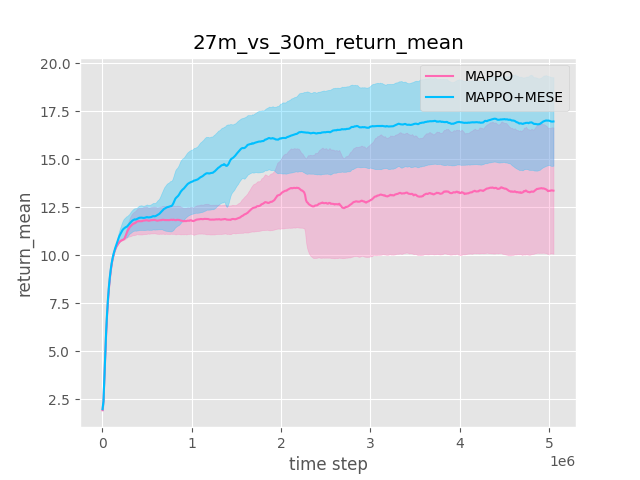}
		\caption{27m vs 30m}
		\label{27m_vs_30m}
	\end{subfigure}
	\centering
	\begin{subfigure}{0.33\linewidth}
		\centering
		\includegraphics[width=0.95\linewidth]{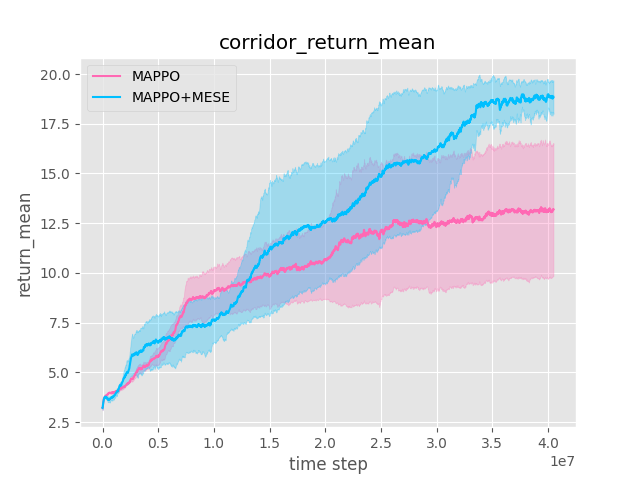}
		\caption{corridor}
		\label{corridor}
	\end{subfigure}
        
	\caption{StarCraft 2 return mean}
 \label{fig_sc2_2}
\end{figure*}

\subsection{Implement}
In this section, we provide a detailed explanation of the \ALGO method used to calculate intrinsic rewards with selected sub-states.

Once a suitable sub-state has been selected in the way mentioned in Section \ref{sec:subspace}, a function is needed to calculate an intrinsic reward that measures sub-state novelty. To this end, we chose the Random Network Distillation (RND) method. This well-known intrinsic reward function uses two neural networks: a fixed and randomly initialized target network $f:\mathbbm{R}^n\rightarrow \mathbbm{R}^p$, and a predictor network $\hat{f}:\mathbbm{R}^n\rightarrow \mathbbm{R}^p$ trained on data collected by agents.

The reward is calculated as the Mean Squared Error (MSE) of the features generated by the target and predictor networks, given by the following equation:
\begin{equation}\label{equ:rnd_loss}
R_{in}(s)=L_{RND}(\phi)=\lVert f_{preditor}(s';\phi)-f_{target}(s')\rVert^2,
\end{equation}
The target network in RND is not trainable, whereas the predictor network is trained using gradient descent to minimize the MSE in Equation (\ref{equ:rnd_loss}). The randomly initialized neural network is distilled into a trained one during the training process. Intuitively, the intrinsic reward is higher for new states and decreases as the predictor network becomes more adept at predicting the target network's features.


In the interest of simplicity, \ALGO employs joint entropy instead of conditional entropy. To achieve this, we apply the chain rule of conditional entropy to rewrite $H(s_j|s_I)$ as follows:
\begin{equation}\label{equ:rnd_loss} H(s_j|s_I)=H(s_j,s_I)-H(s_I). \end{equation}
This form allows \ALGO to estimate the differential entropy of $s_I$ and $s_{I,j}$ directly using the method described in Section \ref{sec:estimator}.

Figure \ref{fig:mrnd} illustrates the most important architectures of the \ALGO. A sketch of the optimization algorithm is presented in Algorithm \ref{alg1}.

\begin{algorithm}
    \renewcommand{\algorithmicrequire}{\textbf{Input:}}
    \renewcommand{\algorithmicrequire}{\textbf{Output:}}
    \caption{\ALGO}
    \label{alg1}
    \begin{algorithmic}[1]
        \STATE Initialization: mask index set $I$, intrinsic reward function $R^{in}_\phi$, MARL algorithm $Agent_\theta $
        \FOR{$e\leftarrow 1,\cdots,E$}{
            \STATE generate episode $(s_1,\boldsymbol{a}_1,r_1,s_2,\dots)$ with $Agent$
            \STATE $\boldsymbol{r}^{\ALGO}=\boldsymbol{r}+R^{in}_\phi(\boldsymbol{s}_{t+1})$
            \STATE train $Agent$ with episode $(s_1,\boldsymbol{a}_1,r_1^{\ALGO},s_2,\dots)$
            \STATE train intrinsic reward function $R^{in}_\phi$
            \IF{$e$ mod $M$ = 0}{
                \STATE update mask index set $I$
            } 
            \ENDIF
                
        }
        \ENDFOR
    \end{algorithmic}
\end{algorithm}

\section{Experiment}
In this section, we present a comprehensive evaluation of the baseline MAPPO algorithm and the modified version of MAPPO that incorporates \ALGO. For our evaluation, we used a benchmark called EPyMARL\cite{epymarl}, which extends the PyMARL\cite{pymarl} codebase to include additional algorithms and allows for flexible configuration of algorithm implementation details. We first evaluate \ALGO on a toy experiment, pushing the box, for considered settings to see whether \ALGO can learn reasonable intrinsic rewards for exploration and comprehensively analyze the algorithm's performance in detail. Then, we study \ALGO in several challenging micromanagement games in StarCraft \Romannum{2}. Both the baseline model and our algorithm were trained using identical hyperparameters, and we conducted five experiments with different random seeds. The solid line in the graph represents the mean performance, while the shaded area represents the standard deviation interval.

\subsection{A Study on Toy Environment}
\begin{figure}
    \centering
    \includegraphics[width=\linewidth]{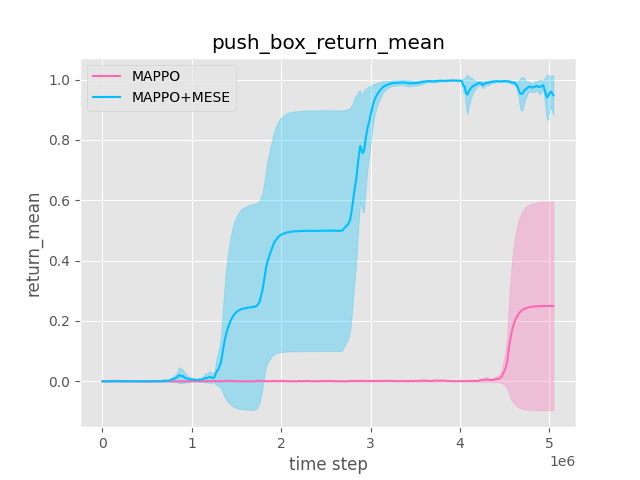}
    \caption{PushBox}
    \label{fig:result_pushbox}
\end{figure}

The pushing box environment depicted in Figure \ref{fig:pushbox} appears deceptively simple but requires considerable exploration. As discussed in Section \ref{sec_intro}, the most critical transition in this environment occurs when (1) all agents are located on the same side of the box and are adjacent to it, and (2) all agents simultaneously move to the opposite side of the box. However, the box coordinates are difficult to modify without a specialized mechanism encouraging exploration. In order to evaluate the effectiveness of \ALGO, we compare the time taken to reach the goal state in this environment using a well-known baseline method, MAPPO, and a modified version of MAPPO that incorporates \ALGO. As shown in Figure \ref{fig:result_pushbox},

\subsection{StarCraft \Romannum{2}}
In this subsection, we evaluate modified MAPPO and baseline in six mini-game. We report the results of six experiments. For each experiment, we report the win rates and returns of the algorithms and provide a detailed analysis of our findings. Our evaluation sheds light on the effectiveness of the modified MAPPO algorithm and its potential to enhance the performance of multi-agent reinforcement learning tasks.

Our experimental results demonstrate that the modified version of MAPPO with \ALGO performs significantly better than the baseline algorithm in several environments, including 3s\_vs\_4z, 3s\_vs\_5z, 27m\_vs\_30m, and corridor. The modified version's enhanced performance can be attributed to \ALGO's effective exploration strategy, which enables the agents to explore the state space more efficiently. This improvement is particularly significant in environments where effective exploration is crucial, such as 3s\_vs\_4z, 3s\_vs\_5z, and corridor.

Interestingly, in the 3s\_vs\_5z environment, the win rates for both the baseline and modified versions are zero. However, the modified version performs better in terms of return than the baseline, suggesting that \ALGO helps the agents make more informative decisions even in the absence of clear win conditions. In the 2c\_vs\_64zg and 3s5z environments, the modified version performs slightly worse than the baseline, but the difference in performance is relatively small.

Overall, our findings suggest that incorporating \ALGO into MAPPO can significantly enhance its performance in certain types of multi-agent reinforcement learning tasks. The success of the modified version highlights the effectiveness of \ALGO's exploration strategy, which has the potential to be applied to other reinforcement learning algorithms.

\subsection{Ablation Study}

To better understand the contribution of \ALGO to the modified MAPPO algorithm's performance, we conducted an ablation study in the pushing box environment. Specifically, we removed the mask from state to sub-state from the \ALGO, i.e., incorporating an original RND into MAPPO. This allows us to compare the performance of the modified MAPPO algorithm with and without the additional search and selection of sub-states.

\begin{figure}[h]
\centering
\includegraphics[width=\linewidth]{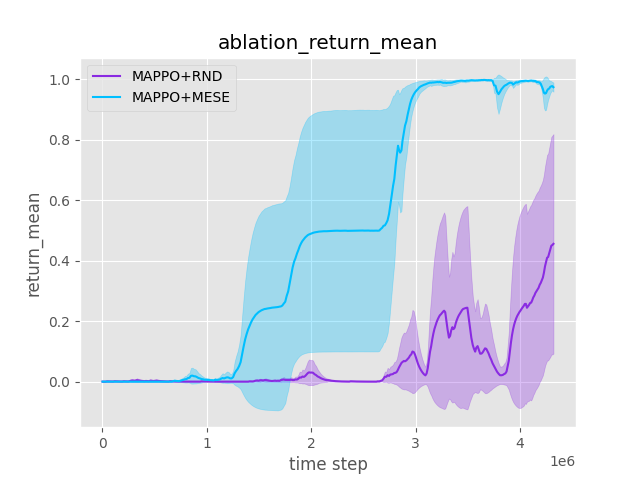}
\caption{Results of the ablation study in the pushing box environment. The modified MAPPO algorithm with \ALGO significantly outperforms the modified MAPPO algorithm with RND.}
\label{fig:ablation_pushbox}
\end{figure}

We trained both versions of the algorithm for 5 million time steps and evaluated their performance on the pushing box environment for 50 episodes. Figure \ref{fig:ablation_pushbox} shows the results of the ablation study. As can be seen, the modified MAPPO algorithm with \ALGO significantly outperforms the modified MAPPO algorithm with RND, achieving a much faster convergence rate and a higher overall reward.

These results demonstrate that the search and selection of sub-states provided by \ALGO is a crucial component of the modified MAPPO algorithm's success in the pushing box environment. Without this additional mechanism for exploration, the modified MAPPO algorithm with RND struggles to learn an effective policy for the pushing box environment. This highlights the importance of \ALGO's ability to identify informative sub-states and use them to guide exploration.

Overall, the results of the ablation study provide further evidence of the effectiveness of \ALGO in enhancing the performance of multi-agent reinforcement learning algorithms. By allowing for more efficient exploration of the state space, \ALGO has the potential to improve the performance of a wide range of MARL tasks.

\section{Conclusion}
Our study proposes \ALGO, a plug-and-play module that improves cooperative exploration in multi-agent reinforcement learning. By incentivizing agents to explore states cooperatively, \ALGO calculates an intrinsic team reward with a sub-state with minimum entropy iteratively, using a particle-based method to estimate high dimensional joint entropy quickly. Our experiments demonstrate that incorporating \ALGO into the MAPPO algorithm significantly improves performance in several challenging environments where effective exploration is critical. Our findings highlight the effectiveness of \ALGO's exploration strategy, which enables agents to explore the state space more efficiently and make more informed decisions. Our results suggest that \ALGO has the potential to enhance the performance of other multi-agent reinforcement learning algorithms.


\clearpage

\bibliography{ecai}

\end{document}